# Einstein y la conjunción entre ciencia, arte y humanidades[*]

Daniel Salomón

CONICET - CeNDIE/ANLIS/Ministerio de Salud, Argentina

1903: Iván Pavlov define en Madrid el reflejo condicionado como el fenómeno psicológico y fisiológico elemental. Cuenta su célebre experimento, "cuando se permite escuchar (al perro operado) el sonido de un metrónomo, la secreción salival comienza luego de nueve segundos". 1976: Umberto Eco escribe en el *Tratado de Semiótica General*, "dos perros se encuentran en Moscú, uno está gordo y bien alimentado, el otro flaco y hambriento. El perro hambriento pregunta al otro: "¿Cómo consigues encontrar comida?", y el otro…responde: "¡Es muy fácil! Cada mañana a mediodía voy al Instituto Pavlov y me pongo a babear, y…, al instante llega un científico condicionado que hace sonar una campanilla y me trae un plato de sopa". ¿Qué ocurrió entre Pavlov y Eco?, entre Pavlov que afirma "los hechos son el aire de los científicos, sin ellos no se puede volar", y Eco que escribe "la semiótica es, en principio, la disciplina que estudia todo lo que puede usarse para mentir. Si una cosa no puede usarse para mentir…tampoco puede usarse para decir la verdad: en realidad, no puede usarse para decir nada." Entre uno y otro han pasado setenta años, pero no setenta años cualquiera, entre el conductista y el semiólogo ha pasado 1905, entre ambos ha ocurrido un Einstein.

Pero a qué Einstein me refiero. No hablaré del físico, excede mi competencia. Hablo del ícono, símbolo, índice, emblema del deseo como de la capacidad de conocimiento. Un Einstein imaginario, extrapolada su relatividad al fin de los absolutos, el dadaísta que proclaman los poetas japoneses tras su gira por la isla en 1922. Un Einstein imaginario e imaginado, pues como dice Todorov "la recepción de los enunciados es más reveladora para la historia de las ideologías que su producción". ¿Qué papel cumple entonces este Einstein en el devenir histórico?, ¿qué papel puede cumplir cualquier individuo en el devenir histórico? Hay quienes proponen que en las reacciones de la historia los individuos somos sustratos simples o productos más o menos complejos, otros suponen que algunos individuos actúan como enzimas y cofactores, compuestos imprescindibles para que las reacciones históricas ocurran, y aún hay quienes suponen que individuos extraordinarios resultan los "primers" o plantillas sobre los que se moldea una época. Considerando todo el espectro de ideologías, desde la dialéctica marxista hasta el existencialismo fenomenológico, opto por la sabiduría desarrollada a partir de milenios de reflexión en las verdulerías, ni muy muy ni tan tan.

Si enuncio: Einstein en 1905 cambia la historia del mundo al publicar, "yo basé mi investigación en las ecuaciones de Maxwell-Hertz para el espacio vacío, junto con la expresión para la energía electromagnética del espacio de Maxwell, y además en el principio…de relatividad". Hay en esa frase una sóla palabra ininteligible para el común de la humanidad, la palabra historia. Eso me obliga a una definición, y evitaré en lo sucesivo caer otra vez en tal pecado, porque con cada definición muere un ángel. En palabras de otro actor de 1905, el pintor Braque, "quien explica (un cuadro) o define algo pone la explicación o la definición en lugar de la cosa. Lo mismo vale para la ciencia. Cada vez que se resuelve un nuevo problema tengo la sensación que se ha

---

perdido algo valioso. En vez de explicar las cosas, sería mejor hacerlas más misteriosas todavía".

Utilicemos, sin embargo, sólo como hipótesis de trabajo, la metáfora de la historia de Fernand Braudel. El devenir humano es un continuo en el tiempo y el espacio pero puede ser enfocado con un microscopio, con lentes de aumentos progresivos. El individuo, su barrio, su ciudad, su país, su civilización, cada uno forma un estrato fluido, moviéndose a diferente velocidad, produciendo interacciones que relacionan el acto personal con los procesos estructurales de larga duración. El pensamiento occidental se puede imaginar entonces como un flujo de partículas individuales, una trayectoria promedio pues la de cada partícula-ideología difiere de las otras y de sí misma en el tiempo. Un movimiento conjunto que de lejos parece homogéneo, el campo de campos entrelazados de la producción simbólica de Bourdieu. Sin embargo ese fluir colectivo del pensamiento en determinados momentos de la historia sufre un punto de inflexión, hace una bisagra y cambia de trayectoria. Las partículas que anticiparon el ángulo de la nueva dirección entonces se catalogan de precursoras, a las que siguen la vieja trayectoria se les cuelga el sanbenito de reaccionarias, y como saben bien los físicos siempre hay partículas con direcciones propias sin sentido, y siempre hay físicos que descomponen los movimientos de las partículas para etiquetarlas como precursoras o reaccionarias.

Así Einstein en 1905 viene a ser el punto de inflexión, emblemático y significativo por sí, de un proceso que podemos ubicar entre 1885 y 1925. Pero un nuevo rumbo sólo se puede comprender a partir de la dirección anterior. Y el cambio de trayectoria previo a 1905 en el pensamiento occidental, permítanme ubicarlo entre los siglos XV y XVII, y su momento emblemático el 12 de octubre de 1492. Luego, qué duda tendría cualquier niño en edad escolar, el equivalente de Einstein viene a ser Piero della Francesca, que falleció el 12 de octubre de 1492. Algunos eruditos reclamarán que en esa fecha también un almirante de origen dudoso se topó con un continente real en busca de uno imaginado. Colón no representa el nuevo pensamiento, queda a horcajadas entre las cruzadas medievales y la curiosidad renacentista, sólo cree en el saber *a priori*, admite "para la ejecución de la empresa de las Indias no me aprovechó razón ni matemática ni mapamundos; llenamente se cumplió lo que dijo Isaías". Sin embargo América sí se constituye en parte del imaginario humanista, fantasía de los exploradores, reciclada en *Utopía* por Tomás Moro, y esa misma *Utopía* devuelta a América en las aldeas de Vasco de Quiroga.

Pero retornemos al 12 de octubre de 1492, Piero della Francesca y la ciudad ideal. Elías y Todorov coinciden en señalar que el estudio de Alberti sobre el arte clásico generó la perspectiva en la pintura, que esta perspectiva generó una observación atenta de la realidad, y esta observación generó la ciencia experimental y el pensamiento moderno. Perspectiva, punto de vista individual, individualidad de las cosas representadas, objetivación de la realidad, secuencia que llevará a Galileo a objetivar el tiempo, ya no un transcurrir sino una variable que se mide, se compara de manera sistemática y repetida. Una alianza promiscua entre ciencia y arte que deriva en la tríada reflexión-observación-experimentación. El pintor Juan Gris, desde el siglo XX juzgará "la perspectiva italiana sólo fue un recurso puesto al servicio de las exigencias científicas de la estética renacentista", implicando así que fue necesaria la ruptura de dicha perspectiva para llegar a la relatividad.

En 1905, derrumbándose el equilibrio de lo ideal, insostenible ya el punto de vista único, cambia la trayectoria de las *ideas direccionadoras* como las llama Hiedegger, provocando un descentramiento del yo en relación al otro, en relación al mundo sensible que lo rodea, y en relación al mundo suprasensible de los conceptos. Tres niveles de descentramiento, como la red de capas históricas, moviéndose a diferente velocidad. Tres niveles que se desarrollan en las cinco dimensiones propuestas por Elías, las tres del espacio, el tiempo, y la de la comunicación entre los hombres. Incorporar esta quinta dimensión del universo, la comunicación interhumana junto al tiempo y al espacio, permite interpretar el saber como parte de un largo proceso de aprendizaje, donde el sujeto de ese saber no es el individuo sino las generaciones.

Tenemos así un poliedro de elementos que interactúan, un proceso con tres niveles de descentramiento, en cinco dimensiones, que se manifiesta en las artes, las ciencias físico-naturales y las ciencias humanas, en sus productos pero también en su raíz compartida. Croce, teórico de la estética, reconoce que arte y ciencia coinciden en tres etapas, en la intuición que las precede, en las metáforas con que se expresan, y en la re-creación del receptor, concluyendo "toda obra de ciencia es a la vez una obra de arte". Monod, bioquímico de la evolución, agrega "todos los hombres de ciencia han debido darse cuenta que su reflexión…no es verbal, es una experiencia imaginaria", y la integra a la dimensión interhumana al afirmar "(hoy) se impone el deber…a los hombres de ciencia (de) considerar a su disciplina dentro del conjunto de la cultura". Por ello, intentaré la aproximación al punto de inflexión 1905 considerando los tres niveles de descentramiento, cinco dimensiones y tres áreas de expresión. Pero antes, coherencia obliga, ubiquemos el contexto.

## COORDENADAS EINSTEIN, 1905

Imperio Austro-Húngaro en el cambio de siglo, indivisible e inseparable proclama su escudo de armas. Musil, en su novela *El hombre sin atributos*, lo describe como el Imperio de Kakania, "un país que por su constitución es liberal, pero su sistema de gobierno es clerical. El sistema de gobierno es clerical, pero la actitud general ante la vida liberal. Donde todos los ciudadanos son iguales ante la ley, pero por supuesto nadie es ciudadano". Autosuficiente centro del mundo, donde sus funcionarios están tan empeñados en festejar los 70 años del imperio, que sólo con asombro descubren que Kakania dejó de existir y se transformó en república. Una clase dirigente que, citando a Bourdieu, inventa "según un arte de inventar ya inventado que… tiende a hacer olvidar que no resuelve más que los problemas que puede proponer o que no propone más que los problemas que puede resolver". El personaje de Musil inspirado en el estadista Rathenau, desde la ironía y el desapego arrogante típicos de la doble monarquía, es incapaz de generar una acción política eficaz para frenar la inhumanidad que avanza; y el Rathenau real, ministro de relaciones exteriores, es asesinado por los proto-nazis en 1922. El imperio indivisible e inseparable se había dividido y separado

Desde una óptica más amplia que la restringida a las revoluciones científicas, Lewis Feuer propone "una teoría de la relatividad requiere más que un Einstein, requiere un clima intelectual, una tensión en el mismo tejido de la sociedad". Es el clima que Marthe Robert define como "sentimiento de impostura de la verdad", "una reacción contra el malestar de Occidente ante sus propios adelantos, que atacan desde todo ángulo a los valores tradicionales sin… sustituirlos, destruirlos de una vez por todas o, simplemente, hacerlos olvidar". Es el vacío vertiginoso de la secularización de la vida y "la muerte de Dios", los conflictos económico-sociales derivados de la aplicación industrial fordista de la ciencia, el apiñamiento de la miseria colectiva en los centros urbanos, el derrumbe del ideal positivista. Entonces llega el tiempo de "los conspiradores de la realidad", los desplazados de los salones oficiales, los desclasados por la academia oficial, las minorías étnicas instruidas que no pueden optar a los privilegios y prestigio de los cenáculos sacralizados, que han roto con sus vínculos de origen sin ser aceptados por la sociedad respetable. Es el tiempo de los que pueden gritar lo que niega el sistema, desafiarlo, pues se sienten libres de obligaciones con una tradición muerta que no representa más su percepción del mundo. Es el tiempo de inventar una nueva forma de inventar. Y el sitio y la época donde se hace más evidente el desmoronamiento del ideario iluminista hasta convertirlo en una burocracia teatral, el sitio y época donde es mayor la concentración de minorías con acceso a la cultura secular pero sin acceso a las academia oficial, es la Europa central entre los siglos XIX y XX.

"A cada época su arte. Al arte su libertad", dice el portal del edificio Secesión de Viena que da el nombre al movimiento de 1897. La Viena de Klimt•, Zweig, Freud, el arquitecto ahistórico Otto Wagner, el pre-Bauhaus Adolf Loos. La Viena de la contracomunidad científica, del grupo Olimpia donde el joven Einstein se reune a discutir los fundamentos de la relatividad con el futuro magnicida del primer ministro de Austria. La Praga de la casa de Brenta Fanta donde se reunen escritores como Rilke y Hašek, pintores como Kupka y Beneš, músicos como Janáček, donde un jueves a la tarde de 1911 podemos encontrar a Einstein y Kafka frente a una cerveza discutiendo sobre el físico filósofo Ernst Mach o las últimas propuestas escandalosas de Freud. La generación de 1905 que expande sus fronteras en el grupo de pintores *El Puente*, en el grupo *El Jinete Azul* de Kandinsky, Marc y Klee; 1905, cuando un crítico escandalizado bautiza al fauvismo al tratar de salvajes a Matisse y Vlaminck, Saussure dicta por primera vez su lingüística general, Weber escribe la *Ética Protestante*, Bateson crea el término genética dándole un contenido que lo proyecta desde la confluencia de Mendel y Darwin hacia la evolución de Dobzhansky, pero también hacia Piaget. Es Europa central, 1905, y el flujo promedio del pensamiento occidental acaba de superar un punto de inflexión, comienza una nueva trayectoria, y el yo ahora queda descentrado frente al otro.

**EL OTRO**
En el centro del mundo medieval se encontraba Dios, los individuos no eran actores sino elementos de una jerarquía divina, lo terrestre apenas un medio para alcanzar el ideal externo e ininteligible. El renacimiento coloca en el centro del universo al yo concreto y racional, que descubre la totalidad de la que forma parte, mientras antes formaba una parte sin un todo físico. Es el yo pienso luego existo cartesiano que sólo puede dar fe de su propia existencia. El hombre retoma su destino desde el individuo, incluso el Calvinismo y su predestinación producen una ética de lo material, "un servicio para dar estructura racional al cosmos" según Weber. Pero el individuo al reconocerse descubre que hay otros, Eco nos dice que el otro nunca nos deja indiferentes, como parte del espacio simbólico del yo su sola presencia desafía nuestra existencia, nuestra certeza ideológica de lo que debe ser el mundo. España en 1492 rechaza al otro interior, moros y judíos, e ignora la humanidad del otro exterior, los amerindios. Y luego de 1492, por cuatro siglos, el otro es relegado a dos categorías, es un igual y debo asimilarlo a mi cultura, o rompo la continuidad del yo y es un diferente, inferior, sin posibilidad de llegar a individuo, a humano completo, por lo que puede ser conquistado, esclavizado, asesinado o comido. Igual o diferente, en ambos casos se identifican los valores propios como objetivos, se identifica al universo con uno mismo. En el pivote de 1905 el individuo-yo, tan cómodo como estaba en el centro del universo es violentamente descentrado, se relativiza frente al otro para quedar en una posición excéntrica. El yo descubre que el otro existe fuera de su yo pienso, entonces uno puede ser un otro para el otro y así observarse a sí mismo como un yo ajeno, pero también descubre que el otro piensa por lo que intentar ser en el otro es la única forma de entender su existencia.

El yo observado como otro comienza a afirmarse con la relatividad del punto de vista, con Freud y el individuo multiplicado en una constelación de yoes fuera del control racional. Al mismo tiempo los pintores expresionistas, el fauvismo, el surrealismo, enseñan una realidad desde el yo interior diferente a la que ve el yo cotidiano. Sartre en el prefacio de *Las Flores del mal* escribe "Baudelaire es el hombre que... se mira para verse mirar", y la filosofía continental explora el ser como vivencia subjetiva en un contexto histórico, el hombre y su circunstancia. Proust en su novela *En busca del tiempo perdido* vivisecciona su yo fluido, multidimensional, sólo comprendido en su totalidad a partir de los detalles. Kafka denuncia la imposible unidad del yo al distribuir entre varios personajes los deseos incompatibles e inconfesables de un único individuo.

La percepción del otro como un yo posible, el ser en el otro, se nutre con el relativismo cultural, la antropología de Boas y Malinowski. La ciencia ya no sólo habla del otro sino con el otro, institucionaliza el conocimiento de la lengua del otro como premisa para comprender su cultura, reconociéndole así la calidad de equivalente. Matisse, Picasso, Kirchner, descubren el arte primitivo entre 1904 y 1906 y lo desprimitivizan. Es el ser en otro que se expresa también en el impresionismo y sus herederos, cuando los pintores incorporan el mundo exterior en el acto creativo. Gauguin comienza a pintar con alegría una muchacha kanaka desnuda, decora el cuadro con flores que parecen fosforescentes, y de pronto la pintura se vuelve sombría, porque "las fosforescencias nocturnas significan para el aborigen que está presente el espíritu de los muertos. Ahora se explica como contenido el terror de la muchacha"; las emociones de la modelo-el otro, le dan significado a una obra que el creador desconocía al crearla, y a la que titula Manao Tupapau en lenguaje nativo. La literatura, por su parte, manifiesta el ser en otro en el fluir de la conciencia, el *Ulises* de Joyce donde autor y lector dejan de ser testigos para ser en el personaje, en su yo incoherente y total. Literatura que luego se realimentará con el cine, cuando se impone la mirada del otro, recortando la escenografía hasta entonces neutral del teatro en fragmentos, aproximaciones y ángulos.

Sin embargo el renacimiento, con su marco de referencia centrado en el individuo que identifica, rechaza o ignora al semejante persiste en el siglo XX tanto en los herederos de la intolerancia, como en los herederos de la revolución iluminista con su "somos todos iguales". Desde 1905 le reconocemos al otro el derecho de ser diferente porque sabemos que uno es el otro del otro, nos ponemos en el lugar del otro; el estudioso de la alteridad Lévinas propone "la superación de sí mediante la epifanía del otro". Postura que evita la banalización del mal descripta por Arendt, pero también evita la banalización del bien, la pérdida de los valores, como los de aquella pintora aficionada del cuento de Salinger que ambiciona fusionar en una misma acuarela a Rembrandt y Walt Disney, sus dos pintores preferidos.

En la película *Mi Tío de América* de Laborit-Resnais, peligrosamente sociobiológica, los actores concluyen que nuestro sistema nervioso crece mediante impresiones de los otros, los otros nos van construyendo. Esta concepción se extendió hasta los semejantes lejanos, el resto de los seres vivos, llevando el problema del otro hasta su límite. Simpson, el evolucionista, plantea que nuestra especie merece ser superior por ser la única que duda si es superior, el filósofo Nagel pregunta *¿Qué se siente ser murciélago?,* se pregunta si es posible ser en un otro inconcebible. Mark Twain, un par de años antes que Einstein, nos cuenta que los insectos científicos deciden hacer una expedición fuera del bosque, llegan por azar a una villa veraniega humana durante el invierno, la recorren y encuentran un museo de curiosidades naturales con una biblioteca. El criptógrafo Piojo de la Madera entonces sacude "a cada alma presente con exaltación y sorpresa", al traducir la frase "en verdad muchos hombres creen que los animales inferiores pueden razonar y hablar entre ellos", frase que en el informe final los insectos comentan: "¡Entonces existen animales inferiores al hombre!". Ahora, como en los dibujos del humorista Gary Larson, el otro lejano nos mira e interpreta.

Un paso más del ser en el otro ya nos lleva al mundo inanimado, a la animación de los objetos, el fetichismo de la mercancía de Marx, la teoría del don, el ser en lo que se dona de Mauss, el intercambio simbólico de Levy-Strauss, o su alienación en lo que se deja de donar para Godelier. Percepción del yo-otro en la que ciencia y arte interactúan para proponer una nueva actitud frente al mundo. Apollinaire dice "hoy los sabios no se atienen ya a las tres dimensiones de la geometría euclidiana. Los pintores han llegado…por intuición, a preocuparse por las nuevas medidas posibles del espacio que, en el lenguaje de los talleres modernos, se designa…con el nombre de cuarta dimensión… manifestación de las aspiraciones…de gran número de jóvenes artistas al contemplar las esculturas egipcias, negras y de Oceanía, al meditar sobre las obras de ciencia". El yo se ha relativizado hasta poder integrarse en la totalidad compleja del exterior sensorial, según las opciones de Kandisky "una calle puede percibirse a través del cristal de una ventana…toda ella semeja un ser vivo del "más allá"…Pero también se puede abrir la puerta: abandonar la soledad, hundirse en el "ser-del-exterior", formar parte de él y sentir sus latidos plenamente". Estamos así frente a un descentramiento en relación al mundo sensible, el segundo nivel de descentramiento provocado por la relativización de los sistemas de referencia.

**EL MUNDO SENSIBLE**
El universo natural del medioevo es un universo dado, como en el verso de Ogden Nash "Dios en su sabiduría hizo la mosca, pero olvidó contarnos para qué". El mundo extenso a partir del siglo XV es el de las leyes racionales, sin causas ocultas, un mundo que puede ser interrogado aunque admite una sola respuesta universal. El progreso desembocará inevitablemente en el ideal, el fin de la historia de Comte y Marx, o el gentleman inglés en el evolucionismo social de Malthus hasta Taylor. El individuo que desalojó a Dios del centro del universo es su propio sistema de referencia, se postula como el observador objetivo de la perspectiva renacentista, donde hay un único punto de vista posible, su ideología o su cultura no tienen contexto espacio-temporal. Entonces llega 1905, Cézanne pinta sus bañistas manejando el color como una dimensión espacial, Rouault, dirá "los artistas subjetivos son tuertos, pero los objetivos son ciegos", y Picasso en 1907, retomando el tema de Cézanne en las Señoritas de Aviñón, firma el epitafio del punto de fuga único, incorpora frente y perfil simultáneos, perspectivas varias y pintura primitiva.

La realidad se ha vuelto múltiple, dinámica, con sistemas de referencia relativos, sólo una realidad entre las muchas posibles y en ocasiones una poco probable. La realidad que se modifica al observarla porque el observador es parte de ella, negando así la existencia del espectador privilegiado, negando la del espectador objetivo al referirlo a su marco histórico social, a su incontrolable constelación psicológica. Se abandona la soberbia de decir cómo es el mundo para proponer una explicación transitoria del mundo.

Feuer destaca que un escritor preferido de Einstein, antes aun de 1905, era Thorstein Veblen defensor del relativismo histórico basado en Marx, quien propone que las leyes económicas son relativas a cada realidad social. Y precisamente en esos años la sociología comienza a construir al sujeto como elemento de una red de relaciones causales y casuales, sincrónicas y diacrónicas, individuales y colectivas, en interacción constante y múltiple con el mundo exterior. Son los años en que Bernard difunde en fisiología el concepto de homeostasis, donde lo aparentemente estático es producto de numerosas fuerzas dinámicas que actúan en diferentes direcciones. Son los años en que la genética comienza su fructífero idilio con la evolución descartando que existan metas definitivas pre-establecidas. Al tiempo que novelas como *Berlín Alexanderplatz* de Döblin, *Contrapunto* de Huxley o *El cuarteto de Alejandría* de Durrell, nos mostrarán luego como una misma realidad cambia a medida que los diferentes personajes se ocupan de ella.

Matisse, con variaciones sobre el tema de Cezánne y Picasso, dice "No puedo jugar con signos que nunca cambian", Eco agrega "lo que entra en crisis es el concepto ingenuo de signo, que se disuelve en un retículo de relaciones múltiples y mutables". Calvino aporta desde la literatura, "en el universo ya no había un continente y un contenido, sino sólo un espesor general de signos superpuestos y aglutinados que ocupaba todo el volumen del espacio…Ya no había modo de establecer un punto de referencia". En 1905 se produce la ruptura con los signos tradicionales, se produce una nueva gramática, la de la ruptura continua, del desajuste continuo. Musil, de formación matemática, describe en su novela al hombre contemporáneo como un *Hombre sin atributos* característicos, porque estos cambian junto a los sistemas de referencia, nos cuenta "para cruzar libremente una puerta abierta, es necesario respetar el hecho que tiene un marco sólido. Este principio…es simplemente un requisito del sentido de realidad. Pero si hay un sentido de realidad, y nadie duda que su existencia se justifica, entonces debe haber también algo que podemos llamar un sentido de posibilidad. Así el sentido de posibilidad puede ser definido como la habilidad para concebir que todo podría ser tal como es, y no dar más importancia a lo que es y a lo que no es". Un escritor aún pleno de certezas como Flaubert, anticipando el fin de la ilusión materialista y del naturalismo se pregunta en *La educación sentimental*, "¿Qué quiere decir eso de la realidad? Unos ven negro, otros azul…Si las cosas siguen así, el arte se convertirá en no sé qué broma pesada inferior a la religión como poesía y a la política como interés." Y precisamente el ejemplo de los colores será tema luego de investigación sobre la relatividad cultural, al descubrir que cada sociedad divide en forma diferente el espectro de lo visible.

El arte, o la "comunicación de complejidades ininteligibles" según Wagensberg, deja de ser arte por un patrón objetivo y pasa a ser arte por la propuesta del creador, la recreación del espectador, y el contexto de creación-recreación. Así el arte en sí mismo y no sólo sus productos se vuelve relativo al sistema de referencia. Por ello quinientos expertos a fin del 2004 declararon al orinal firmado por Duchamp en 1917 como la obra más influyente del arte contemporáneo•. La cosa que deja de ser cosa y sin embargo sigue siéndolo en Duchamp, Man Ray o Jasper Johns. El mundo con tendencia a la simplicidad de Ockham ahora se torna complejo, explota la complejidad de su simplicidad aparente. Kafka utiliza palabras elementales con sentidos múltiples, perro, proceso, condena, despojadas de adjetivos que nos den indicios de su verdadero significado, dice Robert "ninguna imagen aislada (de Kafka) vale por sí misma…(cualquiera) puede ser rectificada por aquella que la precede o que le sigue, y todas en fin son necesarias para hacer aparecer, cierto, no la verdad, sino el poder fetichista de las verdades y contraverdades transmitidas en el lenguaje cotidiano". Recursividad que emparenta las obras de Kafka, Joyce o Proust con el cubismo, con la música serial de Stravinsky, Schönberg, Ravel o Boulez, con la geometría fractal, la parte similar al todo, las dimensiones fraccionarias, ecuaciones de caleidoscopio que retornan a la plástica como arte fractal.

Y si el futuro no puede ser determinado, tampoco lo es el pasado. La historia reconstruida desde el punto de vista único del presente ya no es compatible con la nueva actitud frente al otro y al mundo sensible. Febvre y Bloch, como contrapartida de la geometría no euclidiana incorporan el espacio al tiempo, la geografía a la historia. Pero también se incorpora el otro, Heidegger dice "sólo cuando la calidad de otro de los tiempos pasados se abre paso en la conciencia de un presente, se ha despertado el sentido histórico", un otro cuyo objetivo ya no es ser causa del hoy sino vivir su propio contexto complejo. Confluimos así en Braudel y la historia como capas fluidas, estratos interactuantes de diferente magnitud relativa en el tiempo y el espacio. Confluimos en Elías, la quinta dimensión de la comunicación, la historia privada fruto de una red de relaciones que afecta los hábitos cotidianos, pero a su vez es parte del devenir todo de la humanidad.

Y en ese devenir colectivo artistas y filósofos realimentan sus intuiciones de complejidad transitoria, cuando las ciencias, que se autodenominaban objetivas, afirman que todo punto de vista incluye el contexto del observador. Cuando los investigadores de la materia extensa enseñan que los sólidos son ínfimas porciones de masa entre enormes espacios vacíos, y que aún esa masa es convertible en energía inextensa. Así El Manifiesto de los pintores futuristas en 1910 puede proclamar, "¿Quién puede creer todavía que los cuerpos son opacos?". Las ciencias naturales aportarán a su vez la noción que la estabilidad aparente, desde la escala microscópica a la cósmica, es apenas una construcción cultural, un fotograma de una película continua y azarosa. Las humanidades y el arte se apropian entonces de los términos físicos y biológicos dándole nuevos significados, evolución, ambiente, onda, oscilación, fluctuación. La relatividad ha comenzado a rebalsar el sistema universal supuesto por Einstein, con Mach y Marx a sus flancos, un Einstein que ya en 1928 cauteloso afirma que "principio de covariancia" habría sido una denominación más apropiada para su propuesta que "teoría de la relatividad". Mientras Bohr, con

Berkeley y Kierkegaard montados al hombro, se pregunta si la realidad es una proposición con sentido, y Whitehead afirma que todo objeto tiene una extensión temporal y espacial, que toda la realidad es un proceso, un devenir creativo. Hemos avanzado hacia adelante hasta Heráclito retrocediendo más atrás de Platón, Parménides y sus ideas inmutables. Hemos llegado al tercer nivel del descentramiento, el metafísico.

**EL SER Y EL TIEMPO... Y EL ESPACIO Y LA COMUNICACIÓN**

La Edad Media abordó lo suprasensible como ideas pre-existentes, inmutables, intuidas mediante la revelación. En 1492 Colón se atreve a poner proa al oeste, pero sólo por creer que Asia estaba más cerca de lo que estaba. Alfragan había calculado la circunferencia del globo en millas y el almirante no pudo distinguir entre millas árabes e italianas, no pudo entender que el concepto milla no era universal. Luego de 1905 los conceptos también dependen de las coordenadas de referencia, lo que aún se conserva determinado se torna impredecible. Einstein le dice a Heisenberg: "Usted no debe creer seriamente que nada más que las magnitudes observables deben ir en una teoría física". La relatividad relativiza la misma proposición que la enuncia refiriéndola a un universo dinámico, estadístico, contraintuitivo. La relatividad no es una pérdida de objetividad sino una nueva categoría de objetividad con varios puntos de vista simultáneos. Las leyes son válidas sólo como propuestas que serán indefectiblemente perfeccionadas, las categorías son interpretaciones transitorias que no se anteponen ya a los objetos que categorizan, "las sistematizaciones son *a posteriori*, no marcan líneas futuras" señala Kandinsky. Lo trascendente pasa a ser una hipótesis de trabajo y como tal adquiere el mismo valor que lo cotidiano, lo abstracto es una dimensión más de lo sensible. Llegamos a la concepción del universo cuántico, probabilístico, de la relatividad del significado, donde Murray Gell-Mann toma el término quark de la novela abstracta *Finnegans Wake* de Joyce, Barthes habla del grado cero de la escritura, la literatura explora desde el objetivismo en Robbe-Grillet y Duras hasta los desafíos textuales de Anthony Burgess. Y si la obra de arte revela la verdad de lo ente según Heidegger, entonces el suprematista Malevicht revela una nueva verdad del ente posible, pura, esencial y abstracta.

Se rompe con las viejas formas de expresión, se reagrupan, producen deslizamientos y giros de percepción. El concepto de tiempo continuo y relativo de la física, la psicología de Freud y la filosofía de Bergson se transforman en el fluir de las novelas de Proust, Woolf o Faulkner. Se transforman en el espacio como una dimensión abstracta del tiempo en *El Castillo* de Kafka, que en su *Una confusión cotidiana* escribe, "A tiene que concertar un importante negocio con B, que vive en H. Para las conversaciones previas va a H; hace en diez minutos el camino de ida, y en otros tantos el de vuelta(…). Al día siguiente va otra vez a H (…).Pero aunque todas las circunstancias, por lo menos en opinion de A, son las mismas que en el día anterior, esta vez necesita diez horas para cubrir el trayecto hasta H." El principio de incertidumbre de Heisenberg se ha transformado en final de incertidumbre, el cuento abierto, las novelas sin fin posible de Kafka y Musil.

Demasiado humanos, nada más que humanos, lo suprasensible, aun lo ininteligible, lo puramente intuitivo, lo comunicamos al otro mediante signos. No es casual que 90 años después del *annus mirabilis,* uno de los mayores intentos para enfrentar la desorientación frente a lo fragmentario, para vencer la antinomia entre unicismo y multiplicidad, fue el del congreso interdisciplinario llamado "Einstein encuentra a Magritte", el gran cuestionador del signo pictórico. Y es precisamente el estudio de los signos el que contribuye a relativizar las verdades únicas, demostrando interpretaciones múltiples a partir de un mismo discurso y su contexto. Sapir-Whorf proponen que la visión del mundo de cada cultura está contenida en su estructura gramatical, Wittgenstein reubica al lenguaje en su contexto de construcción cultural, Metz dice que cada palabra es todo un texto, y Hodder "el estudio de todos los textos es hoy una interpretación...No hay una verdad única en un texto o un documento: el sentido es atribuido por la persona que escribe o lee el texto. El estudio de documentos, en consecuencia, no es otro que un análisis crítico".

Y si en las artes verbales la relativización del yo, el mundo físico y el metafísico conducen a explorar el lenguaje mismo, en la plástica y la música alcanzan un inexplorado nivel de abstracción. *Señales abiertas* dice Eco en relación a los cuadros de Mondrian, la música de Schöenberg o post-Weber, imprecisas en el plano de contenido pero con reglas precisas de combinación. Sin embargo, Eco apela al término *galaxias textuales* cuando se refiere a la pintura informal y la música aleatoria, nos dice: "estos nuevos rasgos formales son tan fáciles de reconocer que un ojo medianamente diestro no encuentra la más mínima dificultad para distinguir un cuadro de Pollock de uno de Dubuffet, una composición de Berio de una de Boulez. Señal que hay una organización de nuevas unidades combinatorias: *dicha organización no sigue un código, sino que lo instituye,* y la obra asume también valor metalingüístico". La más mínima característica de significación, de textura, se pone en relieve precisamente para hacer resaltar la ausencia de los fenómenos de significación. Wagensberg dice, cuando la ciencia entra en crisis por el grado de complejidad de la realidad, se utiliza el arte. Por ello aquí entramos por la colectora del arte a la curva 1905 de la autopista hacia lo suprasensible, pero ya en el peaje mismo advertimos que cambiaron las señales, son nuevas y debemos interpretarlas sin manual, sobre la marcha. Si no lo hacemos podemos caer por el primer barranco que mistifique la relatividad, que la entienda como indeterminación absoluta, neutralidad moral. Si no lo hacemos nos podemos estrellar contra el primer animal suelto que se ponga a dar una conferencia como esta.

Einstein en *La física, aventura del pensamiento* acepta que la ciencia se desenvuelve a partir de conceptos, ideas inventadas que crean una imagen transitoria de la realidad. Ideas inventadas como lo fue el continuo unidimensional del tiempo y tridimensional del espacio. Sin embargo Elías previene "cuando los símbolos en el curso de su desarrollo han adquirido un altísimo grado de adecuación con la realidad, los hombres se enfrentan a una dificultad especial para distinguir entre símbolo y realidad". Por ello las ideas inventadas que publica Einstein en 1905 se estrellaron con una visión cristalizada del universo que se había confundido con el universo mismo. Whitehead recuerda el ambiente en la Sociedad Real en Londres cuando el astrónomo real anunció la evidencia que confirmaba las predicciones de Einstein, "el interés y la tensión que llenaban la atmósfera eran exactamente la del drama griego...Había una cualidad dramática en el escenario: la ceremonia tradicional y en el fondo el cuadro de Newton para recordarnos que la mayor de las generalizaciones científicas estaba, ahora, luego de más de dos siglos, por recibir su primera modificación".

Pero no era sólo la primera modificación en mucho tiempo, era una modificación que afectaba a la misma actitud frente al tiempo, que dejó "de ser un punto de referencia externo a la realidad para ser una parte constitutiva de la misma, sin la cual ésta carece de significado", como señalan Toulmin y Goodfield. Arte y ciencia se despojan así del concepto iluminista de tiempo como progreso hacia un fin absoluto, hacia el *estado positivo* "fijo y definitivo...de leyes efectivas...relaciones invariables de sucesión y similitud". Se despojan del concepto de tiempo objetivo, Whitehead podrá decir ahora "en cierto sentido todo está en todo lugar todo el tiempo", porque el proceso de estar siendo y no el ser es el fundamento metafísico del mundo. Un concepto de ser y tiempo más allá de Einstein, pues según Heidegger y Elías el físico sucumbe al fetichismo de las palabras y termina objetivando él también el concepto, habla del tiempo como un objeto extenso, trata de la medición del tiempo y no del tiempo mismo, aunque en su descargo aceptan que tampoco es tarea de la física el hacerlo. Aceptar el concepto de espacio y tiempo objetivo obliga a explicar qué hay antes y después del tiempo, más acá y más allá del espacio. En este sentido una alternativa, tan 1905 como la complejidad, conduce a la unificadora Teoría del Todo, donde la realidad consiste en cuerdas que vibran en diez dimensiones de las cuales percibimos sólo cuatro, reconciliando relatividad general y mecánica cuántica, la escala subatómica y la cósmica. Luego existen universos paralelos en diferentes dimensiones espacio-tiempo, separados por menos de un átomo, invisibles para nuestra percepción, que periódicamente chocan, se aniquilan y reinician el ciclo de nuevo. Una realidad contraintuitiva, que nos deja apenas unos 300 mil millones de años para la próxima colisión y nuevo génesis, una propuesta contraintuitiva como la que le hizo decir a Ernst Mach, el precursor de la relatividad según Einstein, que los átomos son puras abstracciones matemáticas.

Por otra parte, desobjetivar el tiempo y relativizar el ser, ya advertimos, puede llevar al peligroso atajo de la indeterminación absoluta y del nihilismo indiferente. Monod, producto del arte y la ciencia, del renacimiento y la modernidad, descendiente de líderes reformistas, hijo de un pintor, casado con una arqueóloga, premio nobel por sus aportes sobre control genético de la síntesis de las enzimas, escribe "una teoría universal…no podría ser más que estadística... preveería la aparición de objetos como las galaxias o los sistemas planetarios, pero no podría en ningún caso deducir de sus principios la existencia necesaria de tales objetos, de tal acontecimiento, de tal fenómeno particular." Y tomados de la mano de Monod nos ubicamos frente al problema último, los conceptos de absoluto, determinismo y probabilidad que se generan a partir de la inflexión de 1905.

## DETERMINISMO, AZAR y PROBABILIDAD

"¿Acaso puede ocurrir la verdad y ser...histórica?", se pregunta Heidegger. La realidad estadística implica que la percibimos como una realidad entre muchas posibles. Los aportes que la física hizo a este cambio de actitud son indudables, pero su incorporación al pensamiento cotidiano sólo pudo desarrollarse a partir de la interacción entre el arte y la ciencia. Kafka hace de la lógica el argumento de lo fantástico y de lo fantástico un simple accidente de la normalidad. Kandinsky• dice: "El sonido (color) ideal puede ser transformado a través de la asociación con otras formas...(o) cuando la forma aplicada cambia de rumbo. De estas conclusiones se infiere otra: nada hay absoluto."

Rolando García, el decano que sufrió la noche de los bastones largos, comentó recientemente el enojo de René Thom contra quienes "glorifican de manera afrentosa al azar, el ruido, la "fluctuación",...hacen responsable a lo aleatorio, sea de la organización del mundo (vía las estructuras disipativas en el caso de Prigogine), sea de la emergencia de la vida y del pensamiento (vía la síntesis y las mutaciones del ADN en el caso de Monod)". Admitamos con García que una explicación clave para aproximarnos a este mundo probabilístico la dio en 1963 el climatólogo Lorenz, profesión exploradora del azar si las hay, con una publicación de once páginas sentando las bases de la teoría del caos, nombre inadecuado si los hay. Los sistemas ya no son indeterminados pero pequeños cambios en las condiciones iniciales tienen enorme impacto en su devenir, así son determinados pero impredecibles,"no denotan ningún tipo de azar aunque parezcan comportarse al azar". Idea sin embargo compartida por Monod al proponer "el día en que el Zinjántropo, o cualquiera de sus camaradas, usó por vez primera un símbolo articulado para representar una categoría, aumentó por este hecho en proporciones inmensas la probabilidad de que un día emergiera un cerebro capaz de concebir la teoría darwiniana de la evolución"

Una realidad compleja, estadística e indeterminada fue propuesta por físicos como Compton, filósofos como Popper y el matemático padre de la semiótica Charles Pierce. En el otro extremo a partir de la Escuela de Viena, los fisicalistas Neurah y Carnap intentaron ajustar los lenguajes subjetivos al lenguaje objetivo de la física, ignorando "el conocimiento artístico o el conocimiento musical como fuente de la verdad o como foco de revolución" en palabras de Wagensberg, quien agrega "basta que un suceso no sea predecible (por las teorías vigentes) para que el mundo sea indeterminista", "el todo es predecible es enunciable pero no ha sido enunciado". La dificultad de predicción proviene así tanto de la precariedad de las explicaciones, como del azar que determina uno u otro resultado según eventos circunstanciales. Y el azar, determinado o no, resulta "hermoso como el encuentro casual de la máquina de coser con el paraguas en la mesa de disección", en palabras de Isidoro Ducasse, azarosamente uruguayo, casualmente con formación en ciencias exactas, Conde de Lautréamont sin condado, rescatado luego como padre del surrealismo y de los orinales vueltos obras de arte.

**COMPAÑÍA LAMENTABLE**

Hasta aquí he reunido nombres de las artes, ciencias y humanidades en una asociación que, para algunos, puede parecer arbitraria, algo promiscua y decididamente forzada. Y tienen razón, pero no fui el único en señalar un denominador común. Los regímenes reaccionarios reconocen a sus enemigos con más claridad y anticipación que los progresistas. Compañía lamentable, los nazis me precedieron agrupando a muchos de los citados, y cuando Hitler tomó el poder los libros fueron sus primeras víctimas. El 10 de mayo de 1933 encolumnados de antorchas, cantando marchas sobre el coraje y el honor, se quemaron sólo en la Berlin Opernplatz ante 70000 espectadores decenas de miles de ejemplares de los pensadores "degenerados", las obras de Einstein, Brecht, Doblin, Freud, Dos Passos, Kafka, Hemingway, Marx, Hesse, Rilke, Husserl, Kautsky, Mann, Remarque, Trotzky y Wells entre otros. Subhumanos que podían vulnerar a los dioses con el arma de destrucción masiva de un libro, por objetar el universo conservador, por judíos, cosmopolitas o discapacitados como Hellen Keller, un otro diferente inconcebible.

En 1937 son purgados los museos. Se seleccionan 650 obras para la exhibición itinerante de arte degenerado: Chagall, Mondrian, Kandinsky, Dix, Eberhard, Ernst, Klee, Kokoschka, Lange, Klein, el grupo *El Puente*. Siguen luego los conciertos de musica degenerada con piezas de Schönberg, Weill y Mahler. Incluso Nietzsche, Jung y Heidegger, promovidos por el régimen, nunca intentaron ser comprendidos por sus científicos y artistas que se debatían entre el miedo y la fascinación de las certezas. En *Terror y Miseria en el Tercer Reich* Brecht nos muestra a dos físicos nazis haciendo proezas verbales para evitar nombrar la peligrosa E-palabra, Einstein, para evitar nombrar al hombre que los comprende al enunciar "pocas personas son capaces de expresar con ecuanimidad opiniones que difieren de los prejuicios de su medio social."

Bronowsky, el estudioso de los lazos arte-ciencia, dice sobre Auschwitz, "de este charco fluyeron las cenizas de cuatro millones de personas. Y no fue por el gas. Fue por el dogma. Fue por la ignorancia. Cuando la gente cree que tiene el conocimiento absoluto, sin ninguna verificación con la realidad, así es como se comporta". Ante la amenaza de un mundo sin jerarquías, de diferentes pero con iguales derechos, ante lo complejo, la transitorio, lo relativo, hay quienes prefieren refugiarse en el sosiego de los dogmas. Y es este riesgo al dogmatismo el que nos obliga a estar siempre en alerta, repensando nuestras propias ideas desde el otro, el otro en la comunicación, en el espacio y en el tiempo, como advierten Toulmin y Goodfield, "si comienzas por tratar las ideas científicas de los siglos anteriores como mitos, terminarás por tratar vuestras propias ideas científicas como dogmas." Tomemos por ejemplo este pensamiento post-1905: "No hay en el universo ni centro ni circunferencia, sino que el conjunto es central, y también se puede considerar todo punto como parte de una circunferencia, con relación a otro punto central"; esta cita fue escrita por Bruno en *Del universo infinito y los mundos* desafiando el sistema de referencia único, hablando de planetas y otros seres inteligentes posibles, y consecuentemente Giordano Bruno fue quemado vivo el 17 de febrero del año 1600.

El dogma por teleológico implica que los medios siempre se justifican en los fines, por absoluto implica que los que no adhieren a él tienen menos derechos. Los dogmas reducen el caos, el ruido, las fluctuaciones, vuelven al mundo predecible, tranquilizador. Los dogmas, a cambio de certeza sobre el futuro, empobrecen el presente, lo restringen a dialécticas bipolares, impiden la creación. Aceptar la relatividad de nuestras propias proposiciones es una de las contribuciones de Einstein

1905, y no de las menores. Popper destaca que el mismo Einstein definió claramente el dato empírico que podría refutar la teoría general de la relatividad: el no corrimiento al rojo por el potencial gravitacional. Esta actitud no fue sólo honradez intelectual, es parte esencial de la nueva actitud ante el otro, ante la complejidad del mundo, ante la indeterminación de las teorías, ante las leyes vueltas hipótesis de trabajo. Bohr entonces podrá decir a sus estudiantes, "cada frase que yo pronuncio debe ser tomada no como una afirmación sino como una pregunta", Max Born sentenciar "yo estoy convencido que ideas como la certeza absoluta, la precisión absoluta, la verdad final,…son fantasmas que deberían ser excluidos de la ciencia", Heidegger escribir "esta interpretación...es siempre y en todo momento tan correcta y demostrable...que basta para dudar de su verdad", "la esencia de la verdad es la no-verdad", y Picasso refunfuñar "los hombres que quieren explicar un cuadro por lo general ladran en la dirección equivocada. Todos sabemos que el arte no es verdad. El arte es una mentira que nos enseña a comprender la verdad".

El mismo 1905, bisagra de ideas, también debe ser desmitificado. En 1905 Koch recibe el Nobel por sus estudios sobre el cólera y la tuberculosis, hoy aún dos de los mayores fantasmas del mundo sumergido. En 1905 los marineros del Potyomkin se sublevaron contra la comida agusanada, y el siglo que siguió no fue el de la liberación de las clases oprimidas, ni sus demandas fueron satisfechas por el capitalismo ilustrado. En 1905 fue elegido Palacios como primer diputado socialista de América, y la historia argentina en los siguientes cien años no fue precisamente la de una clase política con tolerancia ideológica, honradez, sabiduría e inteligencia. Musil le hace decir a un personaje de su novela, "no hay gran idea que la estupidez no pueda utilizarla para sus propios objetivos", el Talmud, más sentencioso, varios siglos antes proclama "¡Ay de la generación cuyos jueces merecen ser juzgados!". 1905 se estrelló contra 1914, la guerra que pondría fin a todas las guerras. Kafka al leer el primer capítulo de *El Proceso* provocó las carcajadas de sus amigos, el mismo texto que hoy nos aterroriza con sus arrestos y ejecuciones arbitrarios, su policía y leyes secretas. Y debemos reconocer que el cambio del absurdo-denuncia al absurdo del horror y la desigualdad también es siglo XX y reclama nuestra responsabilidad ética.

A poco de andar 1905 dos hechos, fuera de la Gran Guerra, fueron predictivos en el sentido de la teoría del caos: la guerra civil española y el genocidio armenio. A España del 36 no la traigo aquí sólo como ejemplo de la destrucción de un pueblo y su futuro en nombre del dogma, sino también por la incapacidad suicida de las fuerzas progresistas para priorizar objetivos comunes aunque relativos. Al genocidio armenio del 15 no lo traigo aquí sólo por su grado extremo de intolerancia y metódica negación del otro, sino también por la neutralidad moral, el silencio que aún sostiene el occidente democrático por pura especulación política. "¡Abajo la inteligencia!¡Viva la muerte!" gritaron las falanges en la Universidad de Salamanca, ¿seremos capaces de gritar viva la vida siguiendo la inflexión de 1905? El siglo XX, lejos de la irracionalidad fue extremadamente racional, en él compitieron dos actitudes, la inercia positivista transformada en dogma y el pensamiento relativista con sus tres niveles de descentramiento. Uno generó los peores regímenes reaccionarios apelando a ideologías de todo el espectro político. El otro no nos propone soluciones seguras, pero al menos nos ofrece las únicas armas posibles para intentar una sociedad más justa, el conocimiento y la imaginación, el acto creativo. Bronowsky dice "el más hermoso descubrimiento de los científicos es la ciencia en sí misma" y esta vez estoy seguro de no falsear su pensamiento al decir que la más hermosa creación de la ciencia y el arte es la creación en si misma.

Recordemos, sin embargo, que si Galileo y Bruno fueron condenados por la Iglesia, fueron los profesores de filosofía quienes persuadieron al brazo ejecutor, demostrando herejías en el pensamiento nuevo. Recordemos que las listas de arte degenerado de los Nazis fueron escritas por artistas, académicos y estudiantes entusiastas. Recordemos que fue el genetista Lyssenko quien condenó a sus colegas al Gulag, porque su materialismo vulgar no admitía que los genes son responsables de la invariancia y el cambio a la vez. Y a 60 años de los experimentos con humanos de Hiroshima y Nagasaki, recordemos a sus víctimas y recordemos las críticas contra Einstein y Russell en la era post-atómica, cuando Adorno habla de "la estupidez colectiva de los técnicos investigadores", de su "renuncia a pensar", de su autocrítica científica ineficaz por su uso instrumental extracientífico basada en un positivismo idealista. El relativismo en la actitud frente al yo, al mundo sensible y al metafísico, lo transitorio y complejo como premisa, no implican un "dale no más, dale que va". Y si en el cambalache de esta presentación se mezclaron Stravinsky, Don Bosco y La Mignon, el objetivo no fue abrumarlos sino extender el amplio mural de una conjunción, de la confluencia entre arte, ciencias y humanidades que produce el cambio de trayectoria del pensamiento colectivo entre 1885 y 1925. Una confluencia promedio, en la que conviven ideologías y escuelas distintas, a veces circunstancialmente opuestas, pero en la que todos contribuyeron al cambio de trayectoria, todos en algún instante optaron entre el dogma o la relatividad.

Debo comenzar a cerrar esta presentación, y desde los comienzos del arte, desde los comienzos de la ciencia, desde los comienzos de la filosofía, cuando algo no se puede cerrar en su contenido porque es interminable por definición, se cierra limpiamente en lo formal. Comencemos así a cerrar la elipse. En los primeros párrafos utilicé el concepto de historia de Braudel, los estratos temporo-espaciales fluidos, si quieren también relativos, cuánticos, fractales, estadísticos, caóticos, vibrando como cuerdas. Nuestra vida son opciones, opciones que tomamos a cada instante e influyen en nuestro entorno cotidiano, opciones a mediano plazo como colectividad, opciones de larga duración como especie. Observando la corriente de 1905 desde el siglo XXI debemos optar entre los dogmas estáticos o una ética de la tolerancia, entre la tranquilidad de los absolutos o la internalización de la relatividad, entre las certezas de lo dado o el desafío de imaginar un mundo más equitativo aunque complejo, debemos estar preparados para aceptar una nueva actitud frente al universo cuando se agote y dogmatice el pensamiento de 1905. Esta opción no es privativa del último cambio de trayectoria en las ideas, pero lo es su contenido específico. Desde una perspectiva histórica la opción entre dogma y no dogma, entre reacción y progresismo, ha existido siempre y es relativa a su contexto. El desafío a las ideas cristalizadas pudo estar así representado, según la época, alternativamente por el pensamiento paleocristiano y los herejes, por los monárquicos y los anti-monárquicos, por los republicanos, los anarquistas y los socialistas. En el otro extremo del espectro, muchos nacionalistas que hoy defienden las revoluciones patrias, de haber vivido durante esas revoluciones hubiesen defendido con el mismo ardor el *statu quo* colonial.

Sin embargo hoy, los orgullosos herederos de la humanidad que produjo un 1905, cada vez que pensamos como yo debemos pensar también como el otro, debemos vigilar activamente nuestras ideas para evitar la infiltración del dogma y sus dialécticas estrictas, debemos denunciar la aparición del dogma en las ideas circulantes, debemos trabajar para que lo relativo no implique neutralidad ética, debemos crear para un futuro mejor y múltiple contra la desesperación del fragmentarismo y la violencia de la intolerancia. Pues lo fragmentario, ya Comte y la guerra civil española predijeron, conduce hacia atrás, permite la reinvasión del pensamiento mágico hegemónico, del poder que se pretende absoluto. Pues el silencio especulativo frente al dogma, ya el genocidio armenio lo predijo, permite imponer una verdad como única, permite justificar los derechos diferentes de aquellos que califica como diferentes, desde el hambre y la miseria hasta la esclavitud y el asesinato. Y hoy, quienes nos sentimos orgullosos herederos de la humanidad, con la edad de la humanidad a cuestas, seguimos enfrentados a las opciones del dogma frente a la relatividad a cada instante como individuos, hijos-padres, semejantes, miembros de un barrio, de una ciudad, de un país, de un planeta. Pero debo ser coherente con mi adhesión al triple descentramiento, puedo ser en el otro pero nunca opinar por el otro. Así que para concluir en el exacto punto de partida, entre perros, Ivan Pavlov y Umberto Eco, luego de transitar desde el conductismo a la semiótica pasando por 1905, por Einstein y la conjunción de las artes, ciencias y humanidades, por las cinco dimensiones de espacio, el tiempo y la comunicación, por lo relativo, transitorio, complejo y probabilístico, luego de todo este camino que hemos hecho juntos lo único que puedo ofrecerles es mi propia opción. Y en este contexto, mis otras y otros, la única conclusión a mi alcance, la única posibilidad que tengo de transformar este discurso en una acción posible, el contenido específico de mi humilde opción individual hoy y aquí es: para ser el científico de Pavlov prefiero ser el perro de Umberto Eco.